\documentclass[reprint,amsmath,amssymb,aps, twocolumn]{revtex4-2}

\usepackage{hyperref}
\usepackage{comment}
\hypersetup{colorlinks,urlcolor=blue}
\usepackage{graphicx}
\usepackage[percent]{overpic}
\usepackage{dcolumn}
\usepackage{bm}
\usepackage{ulem}
\usepackage{physics}
\usepackage{lipsum}
\usepackage{appendix}
\usepackage{overpic}
\usepackage[T1]{fontenc}

\newcommand{\ham}{\mathcal{H}}
\newcommand{\vect}[1]{\boldsymbol{#1}}

\begin{document}

\preprint{APS/123-QED}
\title{Kinetic magnetism in the crossover between the square and triangular lattice
Fermi-Hubbard models}
\author{Darren Pereira}
\email{dlp263@cornell.edu}
\author{Erich J. Mueller}
\email{em256@cornell.edu}
\affiliation{Laboratory of Atomic and Solid State Physics, Cornell University, Ithaca, New York 14853, USA}

\begin{abstract}
We calculate the spin correlations that result from the motion of a single dopant in the hard-core Fermi-Hubbard model, as the geometry evolves from a square to a triangular lattice. In particular, we consider the square lattice with an additional hopping along one diagonal, whose strength is continuously varied. We use a high-temperature expansion which expresses the partition function as a sum over closed paths taken by the dopant. We sample thousands of diagrams in the space of closed paths using the quantum Monte Carlo approach of Raghavan and Elser [Phys. Rev. Lett. \textbf{75}, 4083 (1995)], which is free of finite-size effects and allows us to simulate temperatures as low as $T \sim 0.3|t|$, even in cases where  there is a sign problem. For the case of a hole dopant, we find a crossover from kinetic ferromagnetism to kinetic antiferromagnetism as the geometry is tuned from square to triangular, which can be observed in current quantum gas microscopes. 
\end{abstract}

\date{\today}
\maketitle

\section{Introduction}
\label{sec:Intro}

Strong short-range interactions can significantly constrain the motion of quantum particles, leading to physics which is very different from what is found in non-interacting systems.  An iconic example is the interaction-driven Mott insulator which is formed when spinful hard-core fermions are placed on the sites of a lattice \cite{HubbardProc1963}.  At a filling of one particle per site, the hard-core constraint prevents the particles from hopping.  Adding a single hole, or a doubly-occupied site (doublon), restores conductivity.  The moving dopant also couples different spin configurations, leading to magnetic order \cite{NagaokaPR1966}. The ordering is ferromagnetic on a bipartite lattice, such as the two-dimensional square lattice \cite{NagaokaPR1966, TasakiPRB1989}. On a non-bipartite lattice, such as the two-dimensional triangular lattice, the type of magnetic correlations depends on the type of dopant:  doublons lead to ferromagnetism, and holes produce antiferromagnetism \cite{HaerterPRL2005}.
Here we use quantum Monte Carlo methods to model the 
kinetically-driven magnetic correlations in the
crossover between square and triangular lattices as a function of temperature.

Due to quantum interference, the energy of a single dopant depends on the state of the surrounding spins.  If the spins are in a completely symmetric, ferromagnetic state, then their only role is to provide a statistical phase when two of them are exchanged.  This statistical phase effectively flips the sign of the hopping for a doublon or hole.  
In a particle-hole symmetric lattice, such as the square lattice, this sign is irrelevant, and the dopant's energy is minimized by this ferromagnetic configuration.  On a particle-hole asymmetric lattice like the triangular lattice, however, a hole in a ferromagnetic background will have a higher kinetic energy than a doublon. This energy can, however, be reduced by adding appropriate spin correlations.  Formally, the ground state of the doublon has maximum spin \cite{NagaokaPR1966}, while the hole's ground state has spin $S=0$ \cite{HaerterPRL2005}.

As a consequence of these energetics, the dopant is dressed by a cloud of magnetic correlations \cite{BohrdtAnnPhys2021, BeranNucPhysB1996, NagaokaPR1966, TrugmanPRB1988}. At finite temperature \cite{BohrdtAnnPhys2021, IgarashiPRB1993, MishchenkoPRB2001, BlomquistCommPhys2020, BlomquistPRR2021, MoreraPRR2023} the spins are uncorrelated far from the dopant, while nearby they are either ferromagnetic or antiferromagnetic.  The size of this so-called polaron grows as temperature is lowered, becoming extensive at $T=0$.  The strength of the correlations also depends on temperature. A highly non-trivial question is how this physics plays out in the crossover between square and triangular lattices, where one expects the correlations to switch from ferromagnetic to antiferromagnetic.

Cold-atom experiments are already able to explore aspects of this physics \cite{BlochNatPhys2012, GreifScience2013, PreissPRA2015,HartNature2015,ParsonsPRL2015, MirandaPRA2015, GreifScience2016, BollScience2016, CheukScience2016, CheukPRL2016, MazurenkoNature2017, GrossScience2017, KoepsellNature2019, YamamotoNJPhys2020, JiPRX2021, GrossNatPhys2021, PrichardNature2024, LebratNature2024, YangPRXQ2021, MongkolkiattichaiPRA2023, BlochNatPhys2005, BlochRMP2008, JordensNature2008, EsslingerAnnuRevCMP2010, BlochNatPhys2012, KoepsellPRL2020, KoepsellScience2020, GreifScience2016, GreifScience2013, HartNature2015, ParsonsPRL2015, ParsonsScience2016, BollScience2016, MazurenkoNature2017, ChiuScience2019, BohrdtNatPhys2019, JiPRX2021, KanaszNagyPRB2017, XuNature2023, LebratNature2024, XuNature2025, ChiuPRL2018, SparPRL2021, GrossNatPhys2021, PrichardNature2024, CheukPRL2016, CheukScience2016, BrownScience2017, TarruellCRP2018, VijayanScience2020, JiPRX2021}.  Fermions are trapped in two-dimensional optical lattices, whose geometry is controllable by manipulating the relative phases of the lattice beams \cite{XuNature2023}.  The experiments are well-modeled by the tight-binding model shown in Fig.~\ref{fig:Hopping}(a).  Sites are defined on a square lattice.  Hopping of strength $t$ is allowed along the cardinal directions, and hopping of strength $t^\prime$ along one of the diagonals.  The case $t^\prime=0$ corresponds to a standard square lattice, while
$t^\prime=t$ is equivalent to a triangular lattice (see Fig.~\ref{fig:Hopping}(b)).  The experiment can continuously vary $t^\prime$ between these limits.  These experiments have single-site resolution: they can simultaneously perform projective measurements of the $z$-component of the spin on every site of the lattice, and the location of every dopant \cite{BlochNatPhys2012, GreifScience2013, PreissPRA2015,HartNature2015,ParsonsPRL2015, MirandaPRA2015, GreifScience2016, BollScience2016, CheukScience2016, CheukPRL2016, MazurenkoNature2017, GrossScience2017, KoepsellNature2019, YamamotoNJPhys2020, JiPRX2021, GrossNatPhys2021, PrichardNature2024, LebratNature2024, YangPRXQ2021, MongkolkiattichaiPRA2023}.  In the limit of low dopant density, they can infer the 
spin-spin correlations in the neighborhood of an individual dopant. Cold-atom experiments have yet to systematically investigate these correlations as one interpolates between the square and triangular lattices. We calculate these correlations while continuously varying this geometry. 

Theoretically, this problem is also relatively unexplored. 
Lisandrini \textit{et al.} used
density-matrix renormalization group methods 
to study this physics \cite{LisandriniPRB2017}, but their work was restricted to the ground state of finite-size systems ($\sim 54$ sites), where they find a continuous phase transition between a ferromagnetic and antiferromagnetic state as $t^\prime$ evolves.  Our study focuses instead on the crossover which occurs at finite temperature. 
Our calculations are free from finite-size effects as
we work in the thermodynamic limit.

The properties of the polaron at the extreme limits of a square or triangular lattice \textit{have} been extensively researched. The early and original works of Nagaoka \cite{NagaokaPR1966} and Thouless \cite{ThoulessProc1965} independently demonstrated that the ground state of the $U=\infty$ Hubbard model for a single hole on a bipartite lattice was the maximal-spin state. Extensions 
then followed to the case of finite $U$ \cite{TasakiPRB1989}, to the $t$--$J$ model \cite{EmeryPRL1990, MartinezPRB1991, LiuPRB1991, OgataPTL1991, LiuPRB1992, GiamarchiPRB1993, WengPRB1995, BeranNucPhysB1996, WhitePRB1997, WengPRB1997, ChernyshevPRB1998, BrunnerPRB2000, MishchenkoPRB2001, Vidmar2013}, and to the problem of a single dopant in a quantum magnet more generally \cite{BrinkmanPRB1970, IordanskiiJETP1980, SchmittRinkPRL1988, ShraimanPRL1988_1, TrugmanPRB1988, ShraimanPRL1988_2, KanePRB1989, SachdevPRB1989, TrugmanPRB1990, SzczepanskiPRB1990, AuerbachPRL1991, BoninsegniPRB1991, BoninsegniPRB1992, MielkeJPhysA1992, IgarashiPRB1993, MielkeCommMatPhys1993, RaghavanThesis, RaghavanPRL1995, HanischAnnPhys1995, StarykhPRB1995, ShengPRL1996, BeccaPRL2001, MerinoPRB2006, ManousakisPRB2007}. A significant extension came from Haerter and Shastry \cite{HaerterPRL2005}, who considered the case of a non-bipartite lattice,  demonstrating that kinetic \textit{antiferromagnetism} arises on the triangular lattice with some nonzero wave-vector. Kinetic antiferromagnetism was later found to occur more generally when kinetic frustration is present \cite{WangPRL2008, SposettiPRL2014, PowellPRB2017, LisandriniPRB2017}. A number of computational techniques \cite{LeBlancPRX2015, QinAnnuRevCMP2022} including 
exact diagonalization 
\cite{DagottoPRB1989, DagottoPRB1990, ElserPRB1990, LeungPRB1995, LeungPRB2002}, 
the density-matrix renormalization group \cite{WhitePRL1992, SchollwockRMP2005, SchollwockAnnPhys2011, WhitePRB1997, WhitePRB2001}, and dynamical mean-field theory \cite{GeorgesRMP1996, ParkPRB2008, ObermeierPRB1997, ZitzlerEPJB2002, SangiovanniPRB2006}
were developed and applied to studying 
the Nagaoka effect and its generalizations.  Some works have looked 
at the crossover between square and triangular lattice geometries, but in the absence of holes \cite{ChangPRB2013}. More recently, the polaron has been explored experimentally using cold atoms trapped in optical lattices \cite{BohrdtAnnPhys2021, BlochRMP2008, BlochNatPhys2005, BlochNatPhys2012, GrossNatPhys2021, EsslingerAnnuRevCMP2010, GrossScience2017}.  These experiments have  
a high degree of  control, including the ability to image every individual site in the lattice. They have explored 
the dynamical formation and propagation of the polaron on a square-lattice N\'eel antiferromagnetic background \cite{JiPRX2021}, the crossover from a Fermi liquid to a metal of polarons with increased doping \cite{KoepsellScience2020}, and the 
spin-spin correlation functions surrounding a particle or hole dopant \cite{KoepsellNature2019, XuNature2023, PrichardNature2024, LebratNature2024}. 
This influx of new experimental data and capabilities has generated a resurgence of interest on the theoretical front. In particular, the properties of the polaron, such as its surrounding spin correlations and dynamics, have been studied in contemporary works using density-matrix renormalization group \cite{BohrdtNewJPhys2020, JiangPRR2020, BohrdtPRL2021, JiangPNAS2021, MoreraPRR2023, SchloemerPRB2024, SamajdarPRA2024}, quantum Monte Carlo methods \cite{CarlstromPRL2016, KanaszNagyPRB2017, GrusdtPRX2018, BlomquistCommPhys2020, BlomquistPRR2021, DiamantisNJPhys2021}, the self-consistent Born approximation \cite{NielsenPRB2021, NielsenPRL2022, VandeKraatsPRB2022, BermesPRB2024}, and other approaches \cite{BohrdtNatPhys2019, GrusdtPRB2019, BohrdtPRB2020, HoPNAS2020, ShenJPCL2024}. It is worth drawing attention to parallel developments in other platforms, such as observing the Nagaoka effect with quantum dots \cite{DehollainNature2020} and accessing Hubbard and polaron physics using moir\'e heterostructures \cite{WuPRL2018, TangNature2020, KennesNatPhys2021, LeePRB2023, DavydovaPRB2023, ZhangScipost2023, CiorciaroNature2023, TaoNatPhys2024}.

In Sec.~\ref{subsec:FHM} we describe the Fermi-Hubbard model. In Sec.~\ref{subsec:HTE} we explain how we calculate its properties in the hard-core limit with a single dopant using a high-temperature expansion. Some technical details of the high-temperature expansion are given in Appendix~\ref{subsec:SFAppendix}.  We introduce Raghavan and Elser's resummation technique \cite{RaghavanThesis, RaghavanPRL1995} to accelerate convergence in Sec.~\ref{subsec:Resummation}, which is further explained in Appendix~\ref{subsec:HairAppendix}.  In Sec.~\ref{subsec:QMC} we describe our Monte Carlo algorithm for sampling the high-temperature expansion, with technical parameters of the simulations given in Appendix~\ref{subsec:QMCAppendix}. In Sec.~\ref{sec:Results} we present our results.  There we characterize the spin correlations in the crossover between a square and triangular lattice, as a function of temperature.  We interpret those results in terms of properties of the polaron.  In Sec.~\ref{sec:Summary} we summarize our findings, and provide an outlook.

\section{Setup} 
\label{sec:Setup}

\subsection{Fermi-Hubbard Model}
\label{subsec:FHM}

\begin{figure}[t!]
    \begin{overpic}[width=\linewidth]{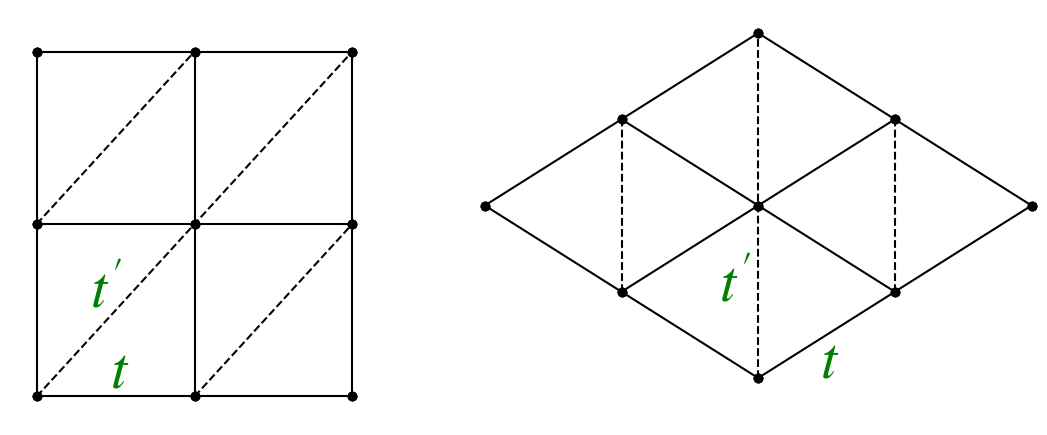}
    \put(-4,40){(a)}
    \put(45,40){(b)}
    \end{overpic}
    \caption{(a) Tight binding model on a square lattice, with hopping $t$ along the cardinal directions and hopping $t^\prime$ along one diagonal. (b) Equivalent depiction of the same model on a triangular lattice, where cardinal hopping $t$ is now along the diagonal directions and diagonal hopping $t^\prime$ is now along the vertical direction.
    This model interpolates between the square lattice ($t^\prime=0$) and the triangular lattice ($t^\prime=t$).}
    \label{fig:Hopping}
\end{figure}

We consider the Fermi-Hubbard model described by 
\begin{align}
    \mathcal{H} = &-t\sum_{\langle ij \rangle, \sigma} \left[c_{i\sigma}^\dagger c_{j\sigma} + \textrm{h.c.}\right] -t'\sum_{\langle\langle ik \rangle\rangle, \sigma} \left[c_{i\sigma}^\dagger c_{k\sigma} + \textrm{h.c.}\right] \nonumber\\& +U\sum_{i} n_{i\uparrow} n_{i\downarrow}, \label{eq:FHM}
\end{align}
where $t$ is the kinetic energy of hopping between nearest neighbors $i$ and $j$ along the edges of the square lattice, $t'$ is the kinetic energy of hopping between next-nearest neighbors $i$ and $k$ along \textit{one} diagonal of the square lattice (see Fig.~\ref{fig:Hopping}), and $U$ is the interaction energy associated with two fermions occupying the same site. Here, $\sigma$ represents the two possible spins of the fermion ($\sigma = \uparrow, \downarrow$). When $t' = 0$, this model represents fermions interacting on the square lattice, whereas when $t' = t$, this model represents fermions interacting on the triangular lattice.   No chemical potential is needed, as we work in the canonical ensemble with a fixed number of particles.

When $U = \infty$ the half-filled ground state, with one particle per site, is highly degenerate.  All spin configurations have the same energy.  Adding a single dopant (either a hole or a particle, which creates an empty or doubly-occupied site, respectively) breaks this degeneracy.  The Hamiltonian $\cal H$ allows the dopant to hop, mixing the spin states and leading to kinetic magnetism.  We solely consider this $U=\infty$ limit, where $\cal H$ 
only consists of dopant hopping terms.  We consider only a single dopant, and all other sites have exactly one particle on them.

In Eq.~\eqref{eq:FHM}, the hopping matrix elements are positive, $t, t' > 0$.
A hole dopant can be mapped onto a doublon via a particle-hole transformation, which flips the sign of both $t$ and $t^\prime$
\cite{FazekasBook}. 
Thus it suffices for us to consider the case of a doublon, taking
$t \to -t$ and $t' \to -t'$ 
when we wish to describe a hole.

\subsection{High-Temperature Expansion}
\label{subsec:HTE}

To calculate correlation functions at finite temperatures $T = 1/\beta$, we use a high-temperature expansion \cite{RaghavanThesis, RaghavanPRL1995, YanayPRA2013, MoreraPRR2023}. Within the canonical ensemble, the partition function on a lattice of $N$ sites can be calculated as
\begin{equation}
    \frac{Z}{\Omega} = \frac{1}{\Omega}\Tr\left[e^{-\beta \ham}\right] = 
    \frac{1}{\Omega}\sum_{x_0,m} \bra{x_0,m} e^{-\beta \ham} \ket{x_0,m},
\end{equation}
where the states $\{\ket{x_0,m}\}$ describe all $N$ possible positions $x_0$ of the dopant and all possible spin configurations $m = \{\sigma_1, \sigma_2, ... \sigma_{N-1}\}$ on the remaining sites of the lattice. Here, $\Omega = N 2^{N-1}$ is the total number of quantum states. By translation symmetry in the thermodynamic limit, all $N \to\infty$ choices of $x_0$ are equivalent. Thus we can fix $x_0$ at the origin and multiply by $N$ to write
\begin{equation}
    \frac{Z}{\Omega} = \frac{N}{\Omega}\sum_{m} \bra{m} e^{-\beta \ham} \ket{m} = \frac{N}{\Omega}\sum_{m} \bra{m} \sum_n \frac{(-\beta \ham)^n}{n!} \ket{m}. \label{eq:HTETrace}
\end{equation}
We can interpret Eq.~(\ref{eq:HTETrace}) as a sum over all paths taken by the dopant.  In particular, the operator $\mathcal{H}^n$ is the sum over all terms where the dopant hops by $n$ sites.  A given path appears exactly once in the expansion.

The expectation value in Eq.~(\ref{eq:HTETrace}) vanishes unless the dopant returns to the origin, allowing us to restrict ourselves to closed paths.  Additionally, moving the dopant through a closed path $\pi$ will permute the spins, $m\to \mathcal{P}_\pi(m)$, but only paths which leave the spin configuration unchanged (i.e. $m=\mathcal{P}_\pi(m)$) contribute. 
A given path will have $\ell_\pi$ hops along the cardinal directions (corresponding to $t$ in $\mathcal{H}$) and $\ell_\pi^\prime$ diagonal hops (corresponding to $t^\prime$).  The partition function is then
\begin{equation}
    \frac{Z}{\Omega} = \sum_\pi \frac{(\beta t)^{\ell_\pi} (\beta t')^{\ell_\pi^\prime}}{(\ell_\pi+\ell_\pi^\prime)!} s_\pi, \label{eq:ZAll}
\end{equation}
where the spin factor $s_\pi$ is defined as
\begin{align}
    s_\pi &= \frac{1}{2^{N-1}} \sum_{m}\langle m| \mathcal{P}_\pi(m)\rangle\label{eq:SpinFactor}.
\end{align}
Here, $\langle m| \mathcal{P}_\pi(m)\rangle$ is unity if $m=\mathcal{P}_\pi(m)$ and otherwise vanishes, enforcing that the underlying spin configuration is unchanged.

Given an operator $X$ which is diagonal in the spin basis,  such as the nearest-neighbor spin correlator $X=S_i^z S_j^z$, we can calculate its thermal expectation value predicated on a dopant at the origin.
 We introduce the projector $h_0^\dagger h_0$, which equals unity if the dopant is at the origin but vanishes otherwise, $h_0^\dagger h_0 |x_0,m\rangle = \delta_{x_0,0}  |x_0,m\rangle$.
  Following the same steps that we used to analyze the partition function, we write the conditioned expectation value as
\begin{align}
    \expval{h_0^\dagger h_0 X} &= \frac{1}{Z} \Tr\left[h_0^\dagger h_0 X e^{-\beta \ham}\right] \nonumber \\
    &=\frac{\Omega}{Z}
    \sum_\pi \frac{(\beta t)^{\ell_\pi} (\beta t')^{\ell_\pi^\prime}}{(\ell_\pi+\ell_\pi^\prime)!} s_\pi(X),
\end{align}
  We have 
defined the \textit{weighted} spin factor
\begin{align}
    s_\pi(X) 
    &= \frac{1}{2^{N-1}} \sum_{m} \bra{m}X \ket{m} \bra{m} \ket{\mathcal{P}_\pi(m)}. \label{eq:WeightedSpinFactor} 
\end{align}
We detail how to calculate the spin factor and weighted spin factor in the thermodynamic limit, $N\to\infty$, in Appendix~\ref{subsec:SFAppendix}.

\subsection{Resummation of Closed Paths}
\label{subsec:Resummation}

\begin{figure}[t!]
    \centering    
    \includegraphics[width=\linewidth]{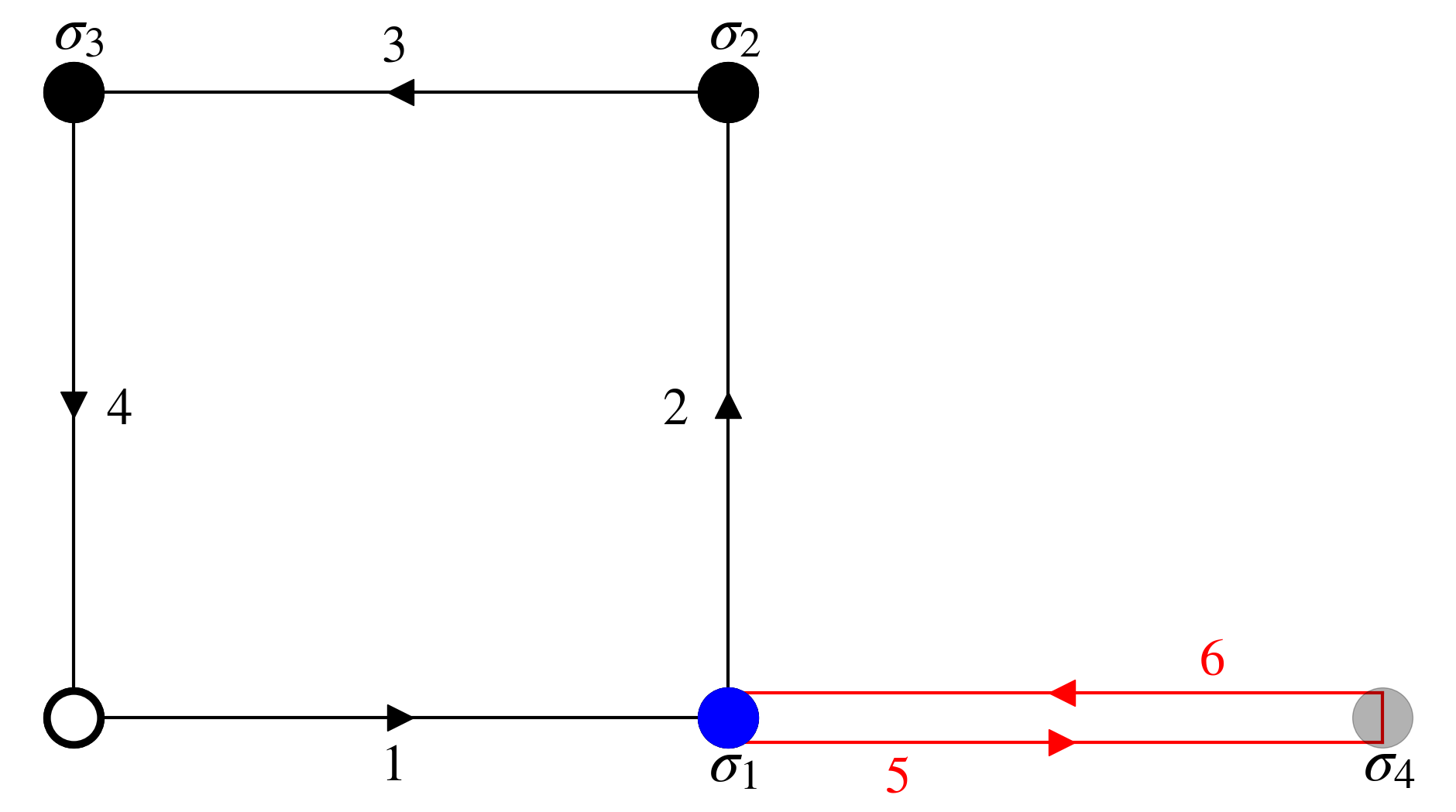}
    \caption{Example of a closed path (black lines), with hair (red lines) emanating from the blue vertex to the gray vertex. The dopant is shown at the origin (open circle) and has no spin. The full path \{1, 5, 6, 2, 3, 4\} and the hairless path \{1, 2, 3, 4\} shuffle the spin configuration $\{\sigma_1, \sigma_2, \sigma_3, \sigma_4\}$ in the same way. Hence, both paths produce identical spin factors (Eq.~\eqref{eq:SpinFactor}), allowing such paths to be factored in the high-temperature expansion (Eq.~\eqref{eq:ZAll}).}
    \label{fig:Hair}
\end{figure}

As argued by Raghavan and Elser \cite{RaghavanPRL1995, RaghavanThesis}, one can accelerate convergence of the series by resumming entire classes of paths that all share the same spin factor.  Using this technique, they were able to calculate properties of the square lattice model down to temperatures almost as low as $T/t\sim 10^{-3}$. The key observation is that no spins are permuted by any self-retracing segment of a path.  Thus the permutation $\mathcal{P}_\pi$ and the spin factor $s_\pi(X)$  are unchanged if one adds or removes such segments. Following Raghavan and Elser's nomenclature, we refer to a self-retracing segment as a ``hair'' and a path with no such segments as ``hairless.''  Figure~\ref{fig:Hair}
shows a simple example.  The segments labeled $\{1,2,3,4\}$ form a hairless path.  Adding the hair labeled $\{5,6\}$ yields a hairy path.

As fully detailed in Appendix~\ref{subsec:HairAppendix}, one can express the partition function as a sum over hairless paths $p$ as
\begin{equation}
    \frac{Z}{\Omega} = \sum_{p} s_{p} W_p . \label{eq:ZFInal}
\end{equation}
The weight function $W_p=W(\ell_p, \ell_{p}', \beta t, \beta t')$ involves the sum over all paths which can be produced by adding hair to $p$.  The only features of $p$ which enter are $\ell_p$ and $\ell^\prime_p$,  the number of cardinal and diagonal steps in the path.  In Appendix~\ref{subsec:HairAppendix} we analytically calculate the generating function for $W_p$, which we use in our simulations.  Expectation values are similarly calculated from the hairless paths as
\begin{equation}
    \frac{Z}{\Omega} \expval{X} = \sum_{p} s_{p}(X) W_p. \label{eq:ExpValFinal}
\end{equation}

\subsection{Quantum Monte Carlo Approach}
\label{subsec:QMC}

\begin{figure}[t!]
    \centering    
    \begin{overpic}[width=\linewidth]{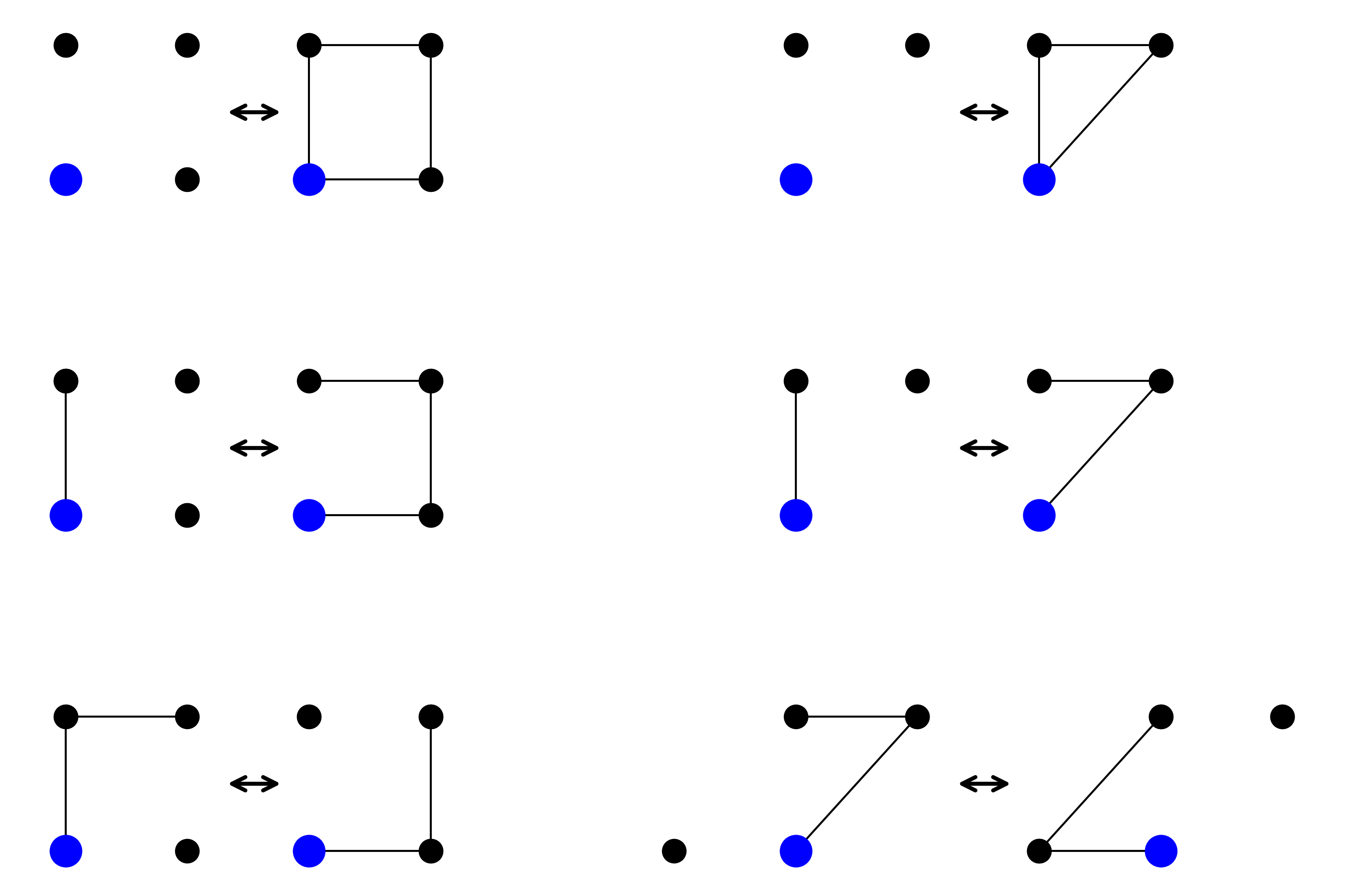}
    \put(-2,64){(a)}
    \put(-2,40){(c)}
    \put(-2,15){(e)}
    \put(52,64){(b)}
    \put(52,40){(d)}
    \put(48,15){(f)}
    \end{overpic}
    \caption{The move set for the quantum Monte Carlo algorithm over the space of hairless closed paths. The left column [(a), (c), (e)] shows moves involving only the cardinal directions with associated hopping $t$; the right column [(b), (d), (f)] shows moves involving the diagonal direction with associated hopping $t^\prime$. There are three classes of moves: plaquette moves [(a), (b)], edge moves [(c), (d)], and corner moves [(e), (f)]. The random point at which the move is being performed is shown in blue.}
    \label{fig:Moves}
\end{figure}

\begin{figure*}[t!]
    \centering    
    \includegraphics[width=\linewidth]{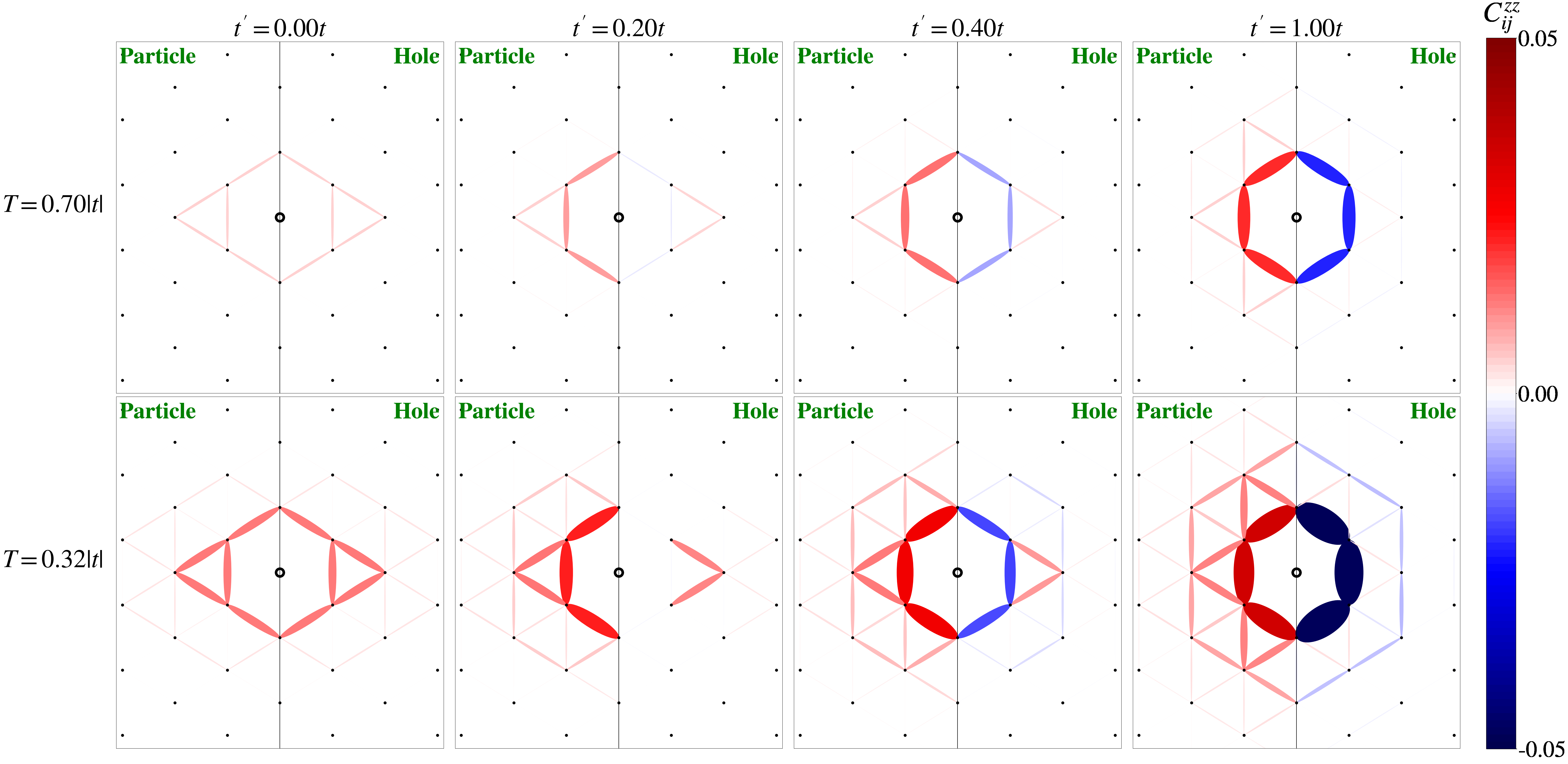}
    \caption{Spin-spin correlation functions, $C_{ij}^{zz} = N\langle h_0^\dagger h_0 S_i^z S_j^z\rangle$, depicted on every nearest-neighbor bond $\langle ij \rangle_{\!_{_\triangle}}\!$ of the triangular lattice for two different choices of the temperature, $T/|t| = \{0.70, 0.32\}$, and four different choices of the diagonal hopping, $t^\prime/t = \{0, 0.2, 0.4, 1\}$ in the thermodynamic limit $N\to \infty$, where $N$ is the number of sites.  
    Here $h_0^\dagger h_0$ is the projector to having a dopant on site $0$, and $C_{ij}^{zz}$ is a three-site correlator involving the position of the dopant and two spins.  It indicates the correlations between two neighboring spins when they are separated from the dopant by some fixed distance.
    See Fig.~\ref{fig:Hopping} for the relationship between the square lattice bonds and the triangular lattice bonds.  The left half of each panel depicts the correlations for a particle dopant, whereas the right half of each panel depicts those for a hole dopant, with the dopant shown at the origin (thick open circle). Red and blue colors indicate ferromagnetic and antiferromagnetic correlations, and the thickness of each ellipse is proportional to $|C_{ij}^{zz}|$. The shade and thickness redundantly indicate the same information, namely the strength of the correlations.  The correlations along square lattice diagonal bonds (which are vertical in this orientation) are always calculated and depicted, although there is no actual hopping along these bonds for the square lattice (i.e. $t^\prime = 0$). The symmetries of the underlying lattice have not been employed to symmetrize the data.}
    \label{fig:NNCorrStrip}
\end{figure*}

The space of all hairless closed paths $p$ is high-dimensional. We therefore perform the sum over $p$ using importance sampling. 
We define a probability distribution,
\begin{equation}
    P_\alpha = \frac{s_\alpha |W_\alpha|}{\sum_p s_p |W_p|}.
\end{equation}
For particle dopants, 
the weight functions are positive and the absolute values can be ignored.
For holes, however, $W_p$ can be either positive or negative.

If we independently sample $M$ paths from this distribution, $\{\alpha_1, \dots, \alpha_M\}$, then we can estimate expectation values of interest as
\begin{equation}
    \expval{X} \approx \frac{\sum_{i = 1}^M (s_{\alpha_i}(X)/s_{\alpha_i})
    \textrm{sign}(W_{\alpha_i})
    }{\sum_{i = 1}^M \textrm{sign}(W_{\alpha_i})}. \label{eq:SampleAvg}
\end{equation}
We use a Markov chain Monte Carlo approach \cite{RaghavanPRL1995, RaghavanThesis, NewmanCompPhys} 
to sample from this space of paths, and estimate such correlation functions.  
Our Markov chain is defined by an ergodic
set of 
transition probabilities $T_{\alpha \to \gamma}$, giving the probability of moving from one state $\alpha$ to the next state $\gamma$ in the chain. These transition probabilities are chosen to satisfy detailed balance,
\begin{equation}
    P_{\alpha} T_{\alpha \to \gamma} = P_{\gamma} T_{\gamma \to \alpha},
\end{equation}
so that it reaches the correct steady-state distribution.  We take $T_{\alpha\to\gamma}$ to be the product of terms which correspond to the probability of proposing a move and accepting it, $T_{\alpha\to\gamma}=P^{\rm prop}_{\alpha\to\gamma} P^{\rm acc}_{\alpha\to\gamma}$.

We consider 
three classes of moves (see Fig.~\ref{fig:Moves}): plaquette moves (which create or destroy a plaquette), edge moves (which extend an edge into a plaquette or vice-versa), and corner moves (which swap the order of two subsequent steps in the path). These are ergodic, as the trivial path can be transformed into an arbitrary path by successively applying these operations.

For the transition probabilities, we use a Metropolis update algorithm. Suppose the hairless path $\alpha$ has $L_\alpha \equiv \ell_\alpha + \ell_\alpha'$ edges. A vertex on this path is selected at random, with $L_\alpha+1$ possible choices (the vertex $x_0$ being counted twice, as the dopant's starting and ending position). We then randomly select a move to perform at the selected vertex.
If that move results in an invalid graph (for example, it adds hair or cannot be performed) the move is rejected.  Otherwise we accept the move with probability
\begin{equation}
    P_{\alpha \to \gamma}^{\rm acc} = \min\left(1, \frac{L_\alpha+1}{L_\gamma+1} \frac{s_\gamma |W_\gamma|}{s_\alpha |W_\alpha|}\right). 
\end{equation}
Here 
${P^{\rm prop}_{\alpha\to\gamma}}{
/P^{\rm prop}_{\gamma\to\alpha}
}=({L_\alpha+1})/({L_\gamma+1})$ 
is the ratio of the  forward and reverse proposal probabilities.
This process of choosing a vertex, choosing a move, and determining acceptance is counted as one sweep. Some additional technical details of the Monte Carlo simulations are given in Appendix~\ref{subsec:QMCAppendix}.

\section{Results}
\label{sec:Results}

Using our algorithm we can calculate any static quantity as a function of temperature and the model parameters.  A particularly relevant quantity is the correlation function $C_{ij}^{zz}=N\langle h_0^\dagger h_0 S_i^z S_j^z\rangle$, which tells us about the alignment of spins which are at positions $i$ and $j$ relative to the location of the dopant, which is placed at the origin.  
This is a special case of the correlator $N\langle h_s^\dagger h_s S_i^z S_j^z\rangle$, where the dopant is at position $s$, and $h_s^\dagger h_s$ is the projector onto the state where a dopant is at site $s$.  By translational symmetry, it only depends on the differences $r_i-r_s$ and $r_j-r_s$.
The normalization factor $N$ corresponds to the the fact that the dopant can be at any of the $N$ sites in the system, which is required for a well-defined thermodynamic limit.
This correlation function has been measured in experiments on square and triangular lattices \cite{LebratNature2024, PrichardNature2024, JiPRX2021}, 
and is studied in several other theoretical works \cite{MoreraPRR2023, KanaszNagyPRB2017}.
By rotational symmetry, we also note $C_{ij}^{zz}=(N/3)\langle h_0^\dagger h_0 {\mathbf{S}_i\cdot \mathbf{S}_j}\rangle$. By construction, this correlator vanishes if $i=0$ or $j=0$, as one cannot have a spin and a dopant on the same site.

We focus on the case where $i$ and $j$ are nearest neighbors on the triangular lattice.  We therefore learn about how short-range magnetic correlations depend on the distance from the dopant. 
As explained in Refs.~\cite{BlochNatPhys2012, GreifScience2013, PreissPRA2015,HartNature2015,ParsonsPRL2015, MirandaPRA2015, GreifScience2016, BollScience2016, CheukScience2016, CheukPRL2016, MazurenkoNature2017, KoepsellNature2019, YamamotoNJPhys2020, JiPRX2021, GrossNatPhys2021, PrichardNature2024, LebratNature2024, YangPRXQ2021, MongkolkiattichaiPRA2023, GrossScience2017}, in a 
quantum gas microscope experiment one extracts this correlation function from
snapshots which show the location of the hole and the spin projections on all sites.  For each nearest-neighbor bond one evaluates, $S_i^z S_j^z=\pm 1/4$, where $i$ and $j$ correspond to the location measured from each hole in the image.  One bins these results, and averages over multiple snapshots.  
In Fig.~\ref{fig:NNCorrStrip} we display $C_{ij}^{zz}$ by drawing ellipses between neighboring sites $i$ and $j$, whose thickness is proportional to the magnitude $|C_{ij}^{zz}|$.  We also color these ellipses so that ferromagnetic correlations are red and antiferromagnetic are blue, and the intensity of the color gives the magnitude.  
There is some redundancy here, as the thickness and shade both indicate the strength of the correlation.
Each panel corresponds to a different value of $t^\prime/t$ and $T/\abs{t}$.  The left half of each panel shows the case of a particle dopant, whereas the right half shows that of a hole.  The sites are displayed in a triangular arrangement, as shown in Fig.~\ref{fig:Hopping}(b).  Thus, when $t^\prime=0$, the square unit cell is displayed as a rhombus. We emphasize that one can still calculate the correlation function along diagonal bonds even when $t^\prime=0$.

\begin{figure*}[t!]
    \centering    
    \begin{overpic}[width=\linewidth]{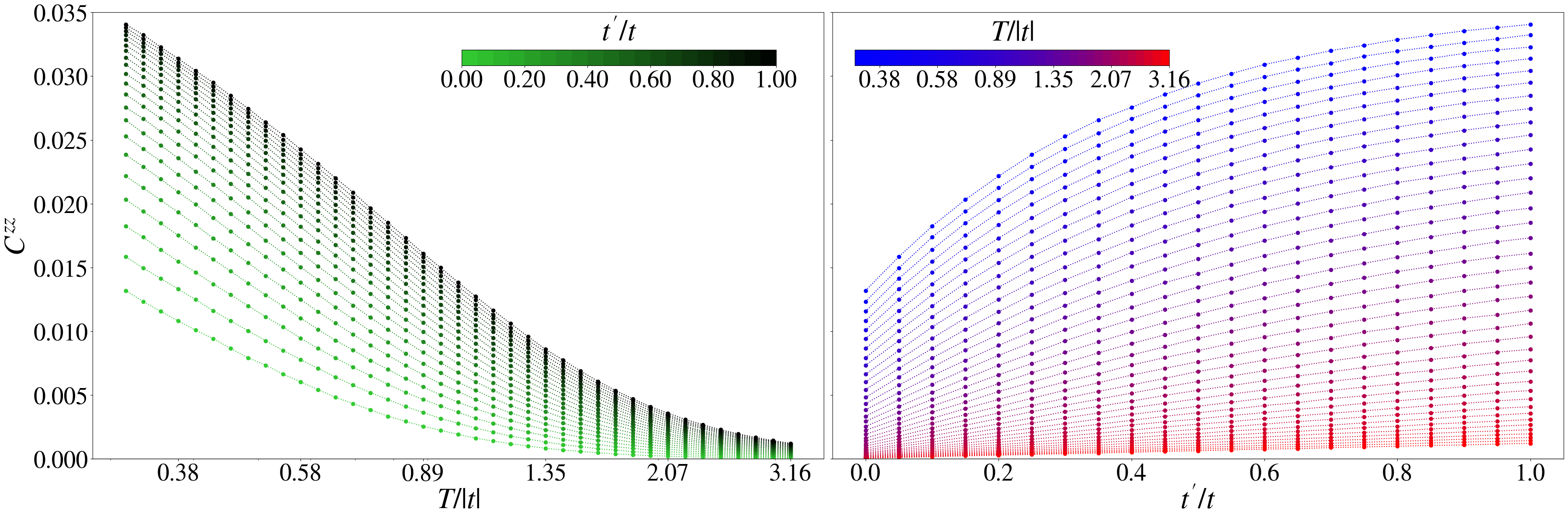}
    \put(-1,34){(a)}
    \put(52,34){(b)}
    \end{overpic}
    \caption{Spin-spin correlation function, $C_{ij}^{zz} = N\langle h_0^\dagger h_0 S_i^z S_j^z\rangle$, for a particle dopant, averaged over the six nearest-neighbor bonds $\langle ij \rangle$ that surround the dopant on the triangular lattice. We denote this averaged correlator as $C^{zz}$.  (a) $C^{zz}$ as a function of temperature $T/|t|$, with color indicating the choice of hopping $t^\prime/t$. (b) $C^{zz}$ as a function of hopping $t^\prime/t$, with color indicating the temperature $T/|t|$.}
    \label{fig:NNCorrParticle}
\end{figure*}

\begin{figure*}[t!]
    \centering    
    \begin{overpic}[width=\linewidth]{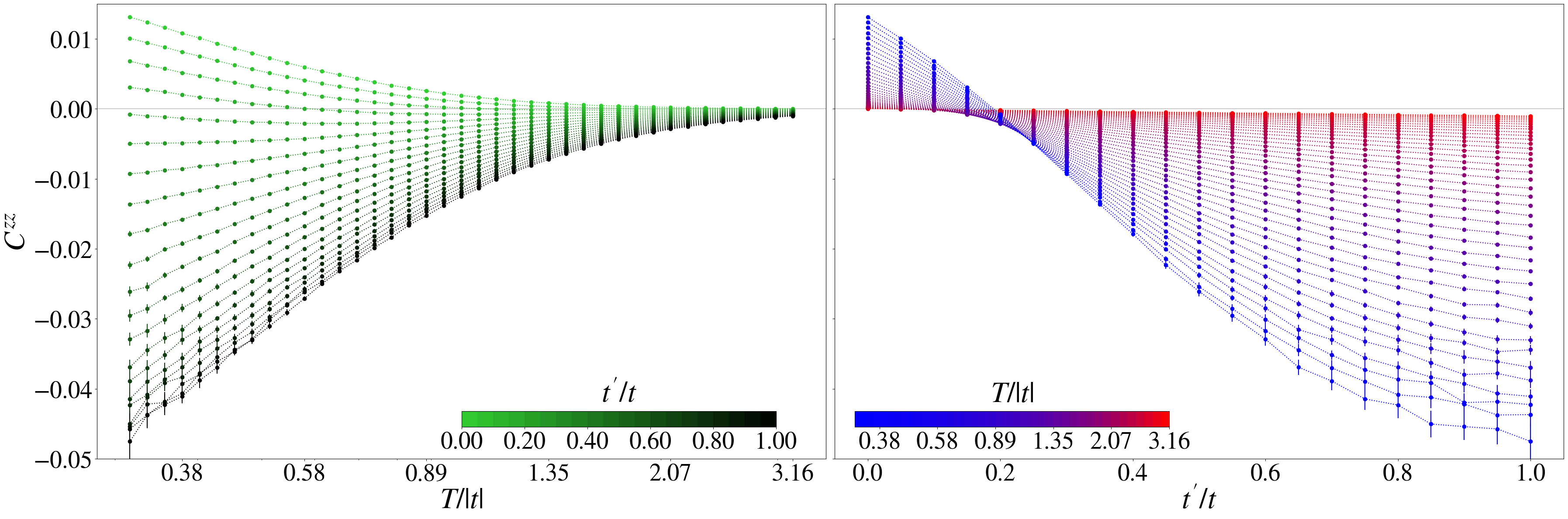}
    \put(-1,34){(a)}
    \put(52,34){(b)}
    \end{overpic}
    \caption{Same as Fig.~\ref{fig:NNCorrParticle}, but for a hole dopant. The gray horizontal line at zero indicates no correlation, with positive and negative values corresponding to ferromagnetic and antiferromagnetic correlations.}
    \label{fig:NNCorrHole}
\end{figure*}

Spin correlations are generally strongest in vicinity of the dopant.  Both the strength and range of the correlations tend to increase with decreasing temperature. The signs of the correlations are sensitive to both the dopant type and the ratio $t^\prime/t$.  Particle dopants  always have ferromagnetic correlations in their vicinity.  Hole dopants can display both ferromagnetic and antiferromagnetic correlations, depending on both $t^\prime$ and $T$.  Particle-hole symmetry is present only at $t^\prime=0$.

It is useful to go through the features seen in each panel of Fig.~\ref{fig:NNCorrStrip}.  The top row corresponds to a moderate temperature, $T=0.70 |t|$.  There are only appreciable correlations on the bonds which are very close to the dopant.  The strength and range of correlations grows with $t^\prime$, and the symmetry evolves from that of the square lattice to that of the triangular lattice.  The most interesting features are for hole doping, corresponding to the right hand side of each panel.  As one increases $t^\prime$, the correlations adjacent to the hole evolve from ferromagnetic to antiferromagnetic.  At $t^\prime=0.2 t$, the correlations on the bonds closest to the hole are very weakly antiferromagnetic, while some of the correlations at larger distances are ferromagnetic and stronger.  These features becomes more pronounced at $t^\prime=0.4 t$, where one sees moderately-strong antiferromagnetic and relatively-weak ferromagentic correlations on different bonds.  As one increases $t^\prime$ further, the ferromagnetic correlations continue to diminish, nearly vanishing at $t^\prime=t$ where antiferromagnetic correlations are most prominent.

At lower temperature, $T=0.32 \abs{t}$, the correlations are almost universally stronger and extend to longer distances.  They follow a similar pattern to the higher temperature data.  Particularly noteworthy is the fact that at $t^\prime=0.2 t$ the correlations essentially vanish on the bonds closest to the hole.  For $t^\prime=0.4 t$, there are interesting structures in the radial dependence of the correlations near the hole.  Antiferromagnetic and ferromagnetic correlations seem to alternate.  This feature persists with increasing $t^\prime$, as shown when $t^\prime = t$. Previous triangle-lattice calculations from Morera \textit{et al.} \cite{MoreraPRR2023} show the same features as our $t^\prime=t$ results.

Experiments at finite hole density on the triangular lattice see similar, but not identical, patterns of correlations \cite{PrichardNature2024, LebratNature2024}.  Unlike our calculations, those experiments observed some antiferromagnetic correlations at larger distances for particle dopants.  Similarly, there are discrepancies in the sign of the correlations at larger distances from hole dopants.  These disagreements could have several sources:  
(1) Our calculations are for a single dopant, while the experiments had finite dopant densities. (2) Our model does not include magnetic interactions due to superexchange.  (3)  There may also be some non-equilibrium considerations in the experiment.

To quantify the patterns and parametric dependence of the spin-spin correlation functions, we average $C_{ij}^{zz}$ over the six nearest-neighbor bonds that form a hexagon around the dopant.  We call this averaged quantity $C^{zz}$.  In Fig.~\ref{fig:NNCorrParticle} we show the temperature and $t^\prime$ dependence of $C^{zz}$ for a particle dopant.  Both panels display the same data.  In Fig.~\ref{fig:NNCorrParticle}(a), the abscissa is $T/|t|$ while each curve corresponds to a different value of $t^\prime$.  Conversely, in Fig.~\ref{fig:NNCorrParticle}(b), the abscissa is $t^\prime/t$.  The temperature scale in Fig.~\ref{fig:NNCorrParticle}(a) is logarithmic.  The behavior shown in this particle-doped case is simple:  lower temperature and larger $t^\prime$ lead to stronger ferromagnetic correlations.  The error bars are small throughout. For the case of the triangular lattice ($t^\prime = t$), our results are in excellent agreement with the Monte Carlo results of Morera \textit{et al.} \cite{MoreraPRR2023}.

We can analytically calculate the expected behavior of $C^{zz}$ as $T \to 0$ and $t^\prime\to 0$, where the ground state is a fully-polarized Nagaoka ferromagnet. Neither our Hamiltonian nor our simulations specify a spin quantization axis, and hence our results correspond to a rotational average over all possibilities.   The ground state density matrix can be written as
$\rho = (2N+1)^{-1} \sum_{m_s}
\ket{m_s}\bra{m_s}$. Here, $\ket{m_s} \equiv\ket{S_{\rm tot} = \frac{N}{2}, S_{\rm tot}^z=m_s}$, where $N$ denotes the number of spinful lattice sites,  $m_s=(N_\uparrow-N_\downarrow)/2$ denotes the magnetic quantum number for the total spin, corresponding to the net spin imbalance, and one sums $m_s$ from $-N/2$ to $N/2$.  
In these fully-polarized states the spin wavefunctions are completely symmetric, and the position of a spin relative to the hole is irrelevant, hence 
$\bra{m_s}S_i^z S_j^z\ket{m_s}=(m_s/N)^2$, as long as $i\neq j$ and $N\gg 1$.    Consequently, 
\begin{align}
\expval{S_i^z S_j^z} &= \frac{1}{2N^3}\sum_{m_s=-N/2}^{N/2} (m_s)^2 = \frac{1}{24}\approx 0.042.
\end{align}
This is a consistent upper bound on Fig.~\ref{fig:NNCorrParticle}. 

The hole-doped case in Fig.~\ref{fig:NNCorrHole} is more structured, as one sees a crossover between ferromagnetic and antiferromagnetic behavior.  In Fig.~\ref{fig:NNCorrHole}(a), the curves fan out as one lowers the temperature, indicating that the $t^\prime$ dependence of the correlations becomes stronger. 
The $t^\prime/t=0.15$ curve crosses the $x$-axis at $T\sim 0.6|t|$, and if one linearly extrapolates the $t^\prime/t=0.2$ curve, it would cross the $x$-axis near $T\sim 0.25|t|$. Such crossings indicate a switch from antiferromagnetic to ferromagnetic behavior.  
Also notable is the fact that these same intermediate $t^\prime$ curves are non-monotonic functions of temperature -- displaying minima.  This property is easier to see in 
Fig.~\ref{fig:NNCorrHole}(b), where the crossing of two curves indicates a non-monotonic temperature dependence.  For $t^\prime>0.3 t$ the curves are monotonic, and the correlations become more antiferromagnetic as temperature is lowered.  For $t^\prime<0.1 t$ they instead become more ferromagnetic.

\begin{figure}[t!]
    \centering    
    \begin{overpic}[width=\linewidth]{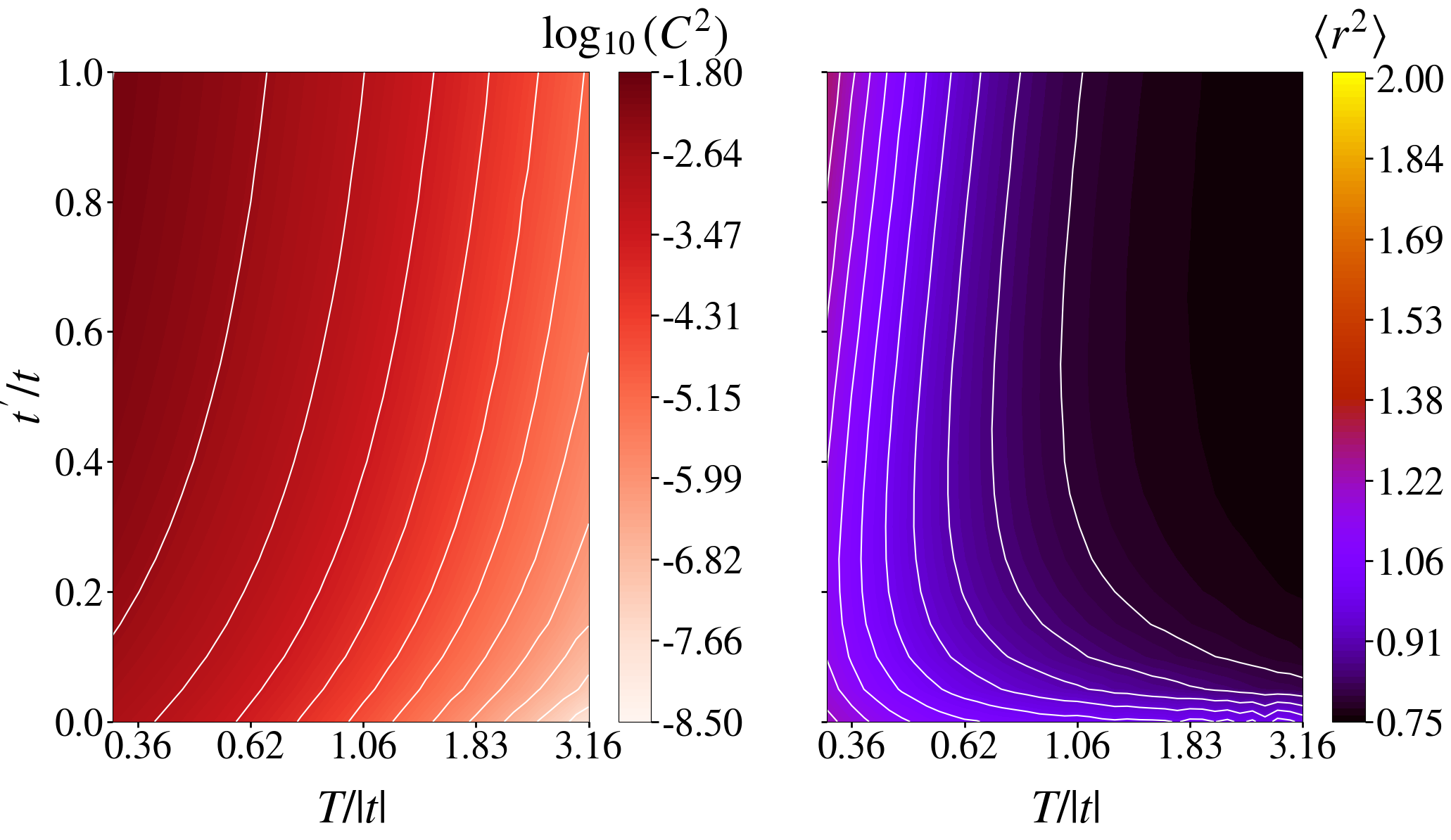}
    \put(-1,56){(a)}
    \put(52,56){(b)}
    \end{overpic}
    \caption{(a) Total magnitude of spin-spin correlation functions, $C^2 = \sum_{\langle ij \rangle} (C_{ij}^{zz})^2$, for a particle dopant. (b) Second moment of the polaron radius, $\expval{r^2} = \frac{1}{C^2} \sum_{\langle ij \rangle} (r_{ij} C_{ij}^{zz})^2$, for a particle dopant.}
    \label{fig:RadiusParticle}
\end{figure}

\begin{figure}[t!]
    \centering    
    \begin{overpic}[width=\linewidth]{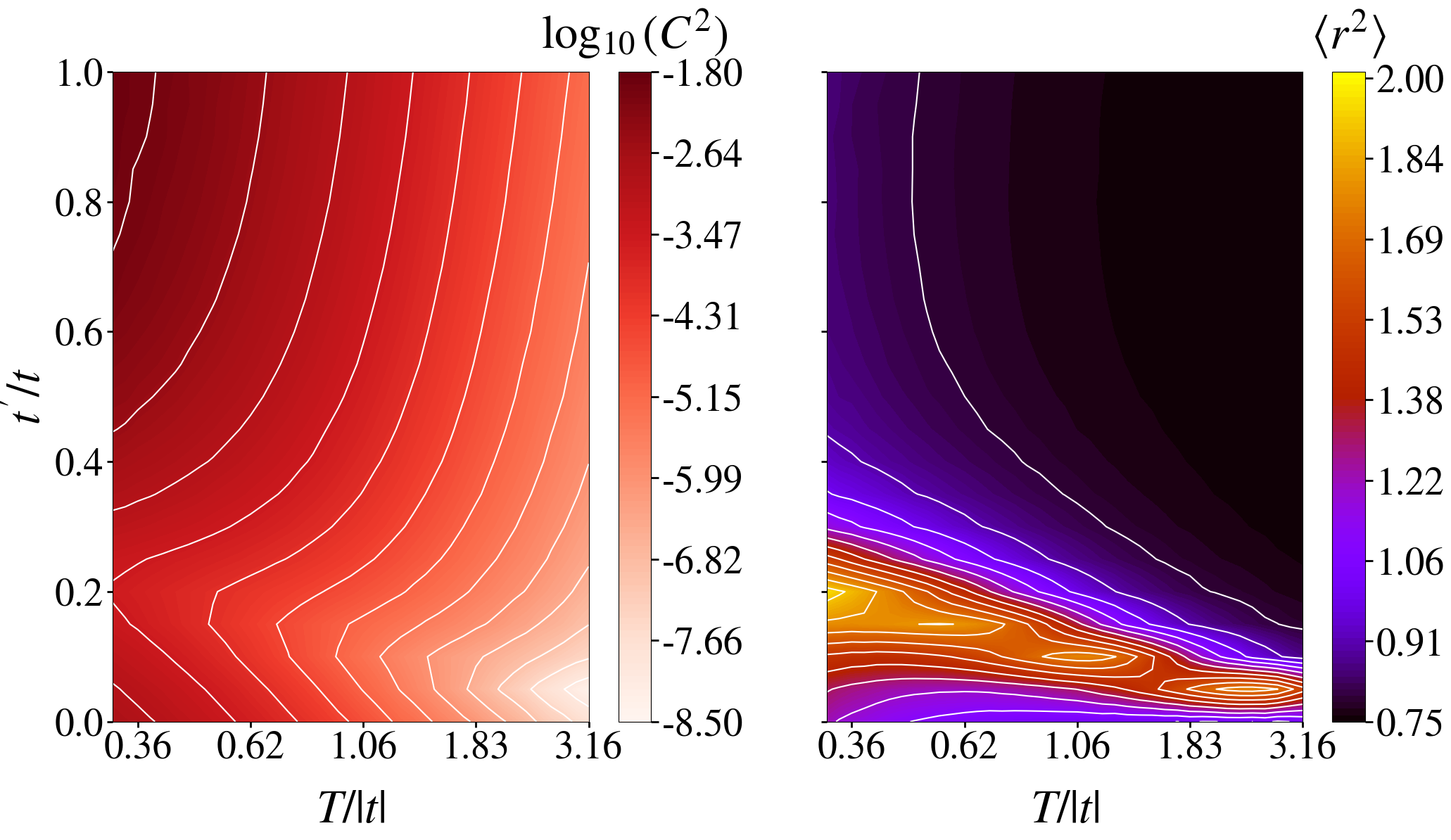}
    \put(-1,56){(a)}
    \put(52,56){(b)}
    \end{overpic}
    \caption{Same as Fig.~\ref{fig:RadiusParticle}, but for a hole dopant.}
    \label{fig:RadiusHole}
\end{figure}

We compare our results to the ground-state
 calculation of 
Lisandrini \textit{et al.} \cite{LisandriniPRB2017}.  They used the density-matrix renormalization group to 
study this model at $T=0$ for a hole dopant. They found a continuous 
phase transition between a uniformly-magnetized state at $t^\prime<t^\prime_{\rm c,GS}\approx 0.22(1) t$, and one in which the magnetization oscillated with some non-zero wave-vector $\vect{Q}$ at $t^\prime>t^\prime_{\rm c,GS}.$
Our results show that at finite temperature the phase transition becomes a crossover.  We  estimate
$t^\prime_{\rm c,GS}$ by extrapolating the crossover location to $T=0$.  
We define $t_{\rm c}^\prime(T)$ as the value of $t^\prime$ where 
the nearest-neighbor correlations $C^{zz}$ vanish. 
We find that $t_{\rm c}^\prime(T)$ is well fit by a sigmoidal in $\log(T)$, and our extrapolation 
suggests $t^\prime_{\rm c,GS} \approx 0.233(1)$. Our estimated uncertainty is purely statistical, and does not include systematic effects from our choice of fitting function.

For the case of the triangular lattice ($t^\prime = t$), our results are generally in good agreement with the Monte Carlo results of Morera \textit{et al.} \cite{MoreraPRR2023}. However, there are some discrepancies below $T/\abs{t} \approx 0.6$, where our calculated correlations near a hole are consistently larger in magnitude. At the lowest temperature simulated in Ref.~\cite{MoreraPRR2023}, $T/\abs{t} = 0.4$, our results are larger in magnitude by around 30\%. 
Given the lack of error bars in that paper, we are unable to say if this discrepancy is statistically significant.

To quantify the longer-range correlations we calculate 
\begin{align}
C^2&=\!
\sum_{\phantom{_{\!_{_\triangle}}}\langle ij\rangle_{\!_{_\triangle}}}\! (C_{ij}^{zz})^2,\\
\langle r^2\rangle &=
\frac{1}{C^2}\!
\sum_{\phantom{_{\!_{_\triangle}}}\langle ij\rangle_{\!_{_\triangle}}}\!  (r_{ij} C_{ij}^{zz})^2.
\end{align}
Here $\langle ij\rangle_{\!_{_\triangle}}$ denotes nearest-neighbor bonds on the triangular lattice.  $C^2$ measures the cumulative magnitude of magnetic correlations produced by the mobile dopant.  The moment $\langle r^2\rangle$ measures the average (squared) distance over which spin correlations extend. Thus these can respectively be interpreted as the strength of the polaronic correlations and the polaron's size.

We plot these quantities in Figs.~\ref{fig:RadiusParticle} and \ref{fig:RadiusHole} for particle and hole dopants.
In the particle case the magnitude of the spin correlations, $C^2$ (Fig.~\ref{fig:RadiusParticle}(a)), increases with decreasing temperature $T$ or increasing diagonal hopping $t^\prime$.  Over the scale shown, $C^2$ changes by roughly seven orders of magnitude.  The size of the polarized region around a particle dopant, $\langle r^2\rangle$ (Fig.~\ref{fig:RadiusParticle}(b)), grows with decreasing temperature, but is a slightly non-monotonic function of $t^\prime$.   Over this entire parameter range $\langle r^2\rangle$ is always notably smaller than two.  As $T\to0$ one expects $\langle r^2\rangle\to\infty$, but one would need to cool to significantly lower temperatures than $T=0.3 |t|$ to see an extended polaron.

The correlations near a hole dopant, as shown in Fig.~\ref{fig:RadiusHole}, show more structure.  Figure~\ref{fig:RadiusHole}(a) shows that the strength of the correlations grows with decreasing $T$, but it is a non-monotonic function of $t^\prime$.  In particular, there is a pronounced minimum for $t^\prime/t \sim 0.1$ -- $0.2$.  The $t^\prime$ of the minimum grows slightly as temperature is lowered.  This minimum is a signature of the crossover from ferromagnetic to antiferromagnetic correlations.  While $C^2$ is not sensitive to the sign of correlations, some spins become uncorrelated in the crossover.  Away from this minimum, the magnitudes of the correlations are similar to the case of a particle dopant.  
Related features are seen in Fig.~\ref{fig:RadiusHole}(b).  The polaron size, $\langle r^2\rangle$, is larger near the locations of the minimum in $C^2$.  The dip in $C^2$  is largely due to a suppression of correlations on bonds close to the dopant, meaning that the remaining correlations are concentrated at distances farther from the dopant.  
For such parameters, the polaron is large $(\langle r^2\rangle$ is peaked) but weak ($C^2$ has a minimum).  Conversely, when approaching the triangular lattice geometry ($t^\prime/t=1$) at low temperatures, the polaron is compact and strong.  Given that $\langle r^2\rangle$ never exceeds 2, these correlations are still short-ranged. Also noteworthy is the fact that $\langle r^2\rangle$ is nearly independent of $T$ and $t^\prime$ for $T>0.5\abs{t}$ and $t^\prime>0.5 t$.

Comparing Fig.~\ref{fig:RadiusParticle}(a) and  Fig.~\ref{fig:RadiusHole}(a), the correlations in the triangular geometry, $t^\prime/t=1$, have comparable strength for particle and hole dopants.  These correlations extend to somewhat larger distances for particle dopants, however, as seen by comparing Fig.~\ref{fig:RadiusParticle}(b) and  Fig.~\ref{fig:RadiusHole}(b).  

\section{Summary and Outlook}
\label{sec:Summary}

We have calculated the spin correlations near a hole or particle (doublon) dopant in the crossover between the triangular- and square-lattice Fermi-Hubbard model with hard-core interactions.  Using a Monte Carlo method adapted from Raghavan and Elser \cite{RaghavanPRL1995, RaghavanThesis}, we calculate these correlations as a function of temperature $T$ and a diagonal hopping $t^\prime$, which allows us to interpolate between the two lattice geometries (see Fig.~\ref{fig:Hopping}). 
We work in the thermodynamic limit, and our results contain no finite-size effects.
As is expected \cite{NagaokaPR1966}, a particle dopant  produces ferromagnetic correlations.  These grow stronger as temperature is lowered or as $t^\prime$ increases.  For a hole dopant, the behavior is more complicated.  
We find purely ferromagnetic correlations when $t^\prime$ is very small compared to the nearest-neighbor hopping $t$.  For larger $t^\prime$ we find antiferromagnetic correlations in the immediate proximity of the hole.  
The crossover occurs when $t^\prime/t \sim 0.1-0.2$, depending on temperature.
At larger distances some of the correlations are ferromagnetic and some antiferromagnetic.  At low temperatures these findings are consistent with finite-size, ground-state simulations performed by Lisandrini \textit{et al.} \cite{LisandriniPRB2017}.

The crossover between ferromagnetic and antiferromagnetic behavior is signaled by a number of features in the resulting spin polaron: in the crossover there is a reduction in the total magnitude of polaronic correlations, and an increase in size of the polaron.  Additionally, in the crossover region, the magnitudes of the correlations closest to the hole are non-monotonic functions of temperature.  

These features are readily observable with current quantum gas microscopes. Recent experiments have already demonstrated how to continuously interpolate between the two lattice geometries \cite{XuNature2023} and have imaged the spin correlations around dopants in the square and triangular lattices \cite{JiPRX2021,KoepsellScience2020, KoepsellNature2019, XuNature2023, PrichardNature2024, LebratNature2024}. The required temperature scale, $T/\abs{t} \lesssim 1$, is  within reach of these experiments, especially considering recent cryogenic developments for Fermi-Hubbard microscopes \cite{XuNature2025}.
As a potential application, one could imagine using the spin correlations near a dopant for local thermometry.  This may be especially useful for probing thermal transport.

Methodologically, our technique 
has a number of advantages when compared to previous approaches.
By resumming classes of diagrams we use fewer resources than prior Monte Carlo calculations \cite{CarlstromPRL2016, KanaszNagyPRB2017, MoreraPRR2023}, and can therefore probe lower temperatures.  Compared to matrix-product-state methods \cite{BohrdtNewJPhys2020, BohrdtPRL2021, MoreraPRR2023, SchloemerPRB2024, SamajdarPRA2024}, we do not suffer from finite-size effects.  Approximate techniques, such as dynamical mean-field theory \cite{GeorgesRMP1996, ParkPRB2008, ObermeierPRB1997, ZitzlerEPJB2002, SangiovanniPRB2006}, are complementary, but often uncontrolled. 
Our technique easily generalizes to other lattice geometries \cite{HanischAnnPhys1995, MielkeJPhysA1992, GlittumNature2025, KimPRB2023, KimPRR2024},  hopping matrix elements \cite{JiangPNAS2021, SposettiPRL2014, LisandriniPRB2017, JiangPRB2024, XuScience2024},
or other spin symmetries, such as SU($N$) \cite{KimPRB2023, KimPRR2024, GarciaPadillaPRA2023, GarciaPadillaJPCM2024, SchloemerPRB2024, BotzungPRB2024, CazalillaRep2014, MukherjeeNJPhys2025}. {It could also be used to 
model spin-polaron physics in moir\'e heterostructures
and other condensed matter systems \cite{WuPRL2018, TangNature2020, KennesNatPhys2021, LeePRB2023, DavydovaPRB2023, ZhangScipost2023, CiorciaroNature2023, TaoNatPhys2024}.} 

We work in the hard-core limit $U=\infty$, where all magnetic effects are due to the motion of the dopant.  At finite $U$, superexchange can also lead to magnetic ordering \cite{FazekasBook}.  Superexchange is always antiferromagnetic for fermionic atoms, and either enhances or competes with the kinetic effects, depending on the type of dopant.

There are three principle limitations of our method.  First, it explicitly considers the case of a single dopant, and is not readily generalized to finite doping density.  Second, with hole dopants and $t^\prime\neq 0$, our Monte Carlo summation involves both positive and negative elements, leading to large variance at low temperatures (i.e. it suffers from a ``sign problem'' \cite{Pan2024}).  Third, our model does not contain superexchange, and there is no simple way to add it to the calculation.  In future work we hope to overcome or mitigate some of these challenges.

\section*{acknowledgement}
We are especially grateful to Veit Elser and Ragu Raghavan for numerous helpful exchanges. This material is based upon work supported by the National Science Foundation under Grant No.  PHY-2409403. We also acknowledge the support of the Natural Sciences and Engineering Research Council of Canada (NSERC) (Ref. No. PGSD-567963-2022).


\appendix

\section{Calculation of the Spin Factor}
\label{subsec:SFAppendix}

In this Appendix, we detail how to calculate the spin factor, Eq.~\eqref{eq:SpinFactor}, and weighted spin factor, Eq.~\eqref{eq:WeightedSpinFactor}, for an arbitrary closed path.
As a concrete example, consider the hairless closed path in Fig.~\ref{fig:Hair}. There are three spinful sites, with spins $\sigma_1$, $\sigma_2$, and $\sigma_3$, leading to $2^3=8$ degenerate spin configurations. After the dopant has hopped around the square plaquette 
the three spins will have been permuted: $\sigma_1 \to \sigma_3$, $\sigma_2 \to \sigma_1$, $\sigma_3 \to \sigma_2$. 
The spin factor counts the fraction of spin configurations which are invariant under the permutation.  There are only two such configurations here: $+++$ and $---$, yielding a spin factor $s=\frac{2}{8} = \frac{1}{4}$ for this particular path.

Generically, a closed path $\pi$ involving $N-1$ spinful sites will have $2^{N-1}$ 
spin configurations, $m =  \{\sigma_1, \dots, \sigma_{N-1}\}$.
These will be permuted to $P_\pi(m)$.  $P_\pi$ can generally be decomposed into $N_{\rm c}$ cycles.  Within one cycle, the spins must all be aligned in order to contribute to the spin factor, and hence $s_\pi=2^{N_{\rm c}}/2^{N-1}$.  Note, the only relevant feature of the path is the number of disjoint cycles in the permutation.  This is very different from a Monte Carlo approach where one directly samples spin distributions.

We now turn to the weighted spin factor, Eq.~\eqref{eq:WeightedSpinFactor}, for  $X=\prod_{j\in A} S^z_j$, which is the product of the spin operators $S^z$ on  $n$ sites in some set $A$.  For example, $A$ may consist of two nearest-neighbor sites -- yielding the spin correlations in Fig.~\ref{fig:NNCorrStrip}.
We will assume $n$ is even, for otherwise $\langle X\rangle=0$ by symmetry.
For a given path $\pi$, the weighted spin factor 
$s_\pi(X)$ vanishes unless all of the sites in $A$ lie in the same cycle.
In that case,
$s_\pi(X)= (\frac{1}{2})^n s_\pi$.

\section{Calculation of the Weight Function}
\label{subsec:HairAppendix}

In this Appendix, we demonstrate how to enumerate all possible combinations of self-retracing paths, or ``hairs,'' and calculate their contributions to the weight function $W_p$, which appears in the expressions for the partition function and expectation values, Eqs.~\eqref{eq:ZFInal}-\eqref{eq:ExpValFinal}.

The weight function $W_p$ is given by a sum over all paths $\pi$ which can be constructed by adding hair to the hairless path $p$.  We write this condition as $\pi\supseteq p$, and formally express
\begin{align}
    W_p&\equiv\sum_{\pi\supseteq p}  \frac{(\beta t)^{\ell_\pi} (\beta t')^{\ell_\pi'}}{(\ell_\pi + \ell_\pi')!} \\ \nonumber
    &=
    \sum_{n_1, n_2=0}^{\infty} \frac{(\beta t)^{\ell_p+2n_1} (\beta t')^{\ell_{p}'+2n_2}}{(\ell_p + \ell_{p}' + 2n_1 + 2n_2)!} w(\ell_p, \ell_{p}',n_1,n_2).
\end{align}
Here, $2n_1$ and $2n_2$ are the total number of steps in the hair that have hopping energy $t$ and $t'$.  In this Appendix we term these as ``type-1'' and ``type-2'' steps. Hairs always involve an even number of steps of each type since they are self-retracing. Practically, this means that the sign of the hopping energies does not matter; the tabulation of hairs is equivalent for both particle and hole dopants. In this equation we have introduced $w(\ell_p , \ell_{p}', n_1, n_2)$, the number of ways there are to add such hairs  to our hairless path. As the notation implies, $w$ only depends on $n_1, n_2$, and the number of type-1 and type-2 steps in the hairless path, which we denote $\ell_p, \ell_{p}'$.
Here we derive this result and construct a generating function for $w$.

To that end we first construct some auxiliary functions.  We define $c_{n_1 n_2}^{(i)}$ ($i=1,2$) to be the number of ways of adding a hair at a given vertex with the conditions that: (1) the first step is fixed to be of type $i$;  (2) the hair never fully retraces itself until the very end; (3) the hair has total length $2(n_1 + n_2)$, with a total of $2n_1$ steps of type 1 and $2n_2$ steps of type 2. By definition, this provides a few base cases for our recursive approach: $c_{1,0}^{(1)} = 1$, $c_{0,1}^{(2)} = 1$, and $c_{0,n}^{(1)} = c_{n,0}^{(2)} = 0$ for any choice of $n$. 

Begin by considering $c^{(1)}_{n_1n_2}$, in which there are $c_{1,0}^{(1)}=1$ choices for the first step from the origin to the next vertex.  Subsequently the path will retrace itself to this second vertex $l$ times, where $l$ runs from $1$ to $n_1+n_2-1$.  For each of these subhairs, labelled by $k=1,2,\dots, l$, we can choose the type of the first step, $\nu_k=\{1,2\}$, and the total number of type-1 and type-2 steps, $2i_k$ and $2 j_k$.  We must constrain $\sum_k i_k=n_1-1$ and $\sum_k j_k=n_2$.  If $\nu_k=1$, there are $\gamma_1-1$ ways of choosing the first step of the subhair, where $\gamma_j$ is the coordination number for type-$j$ paths.  We subtract one, as the subhair cannot start out by stepping backwards to the origin.  Consequently there are $(\gamma_1-1)c_{i_k j_k}^{(1)}$ total such subhairs.  Conversely, if $\nu_k=2$ there are $\gamma_2$ ways of choosing the first step, and $\gamma_2 c_{i_kj_k}^{(2)}$ total such subhairs.  We can therefore write 
\begin{align}
    c_{n_1n_2}^{(1)} &= c_{1,0}^{(1)} \sum_{l=1}^{n_1+n_2-1} 
    \prod_{k=1}^l \sum_{ i_k, j_k}
    \sum_{\nu_k=1,2}
    (\gamma_{\nu_k}-\delta_{\nu_k,1})
    c_{i_k j_k}^{(\nu_k)}, 
\end{align}
where, as already stated, $\sum_k i_k=n_1-1$ and $\sum_k j_k=n_2$.  
We can further simplify this notation if we introduce introduce $\overline{\gamma}_1 = \gamma_1 - 1$ and $\overline{\gamma}_2 = \gamma_2 - 1$. Let $s$ be the number of subhairs which start with a type-1 step. Then we can write
\begin{align}
    c_{n_1n_2}^{(1)} &= c_{1,0}^{(1)} \left[\sum_{l=1}^{n_1+n_2-1} \sum_{s=0}^{l} (\overline{\gamma}_1)^s  (\gamma_2)^{l-s} \prod_{k=1}^l \sum_{\substack{\nu_k,\\ i_k, j_k}} c_{i_k j_k}^{(\nu_k)} \right]. \label{eq:C1Coeff}
\end{align}
Here the $\nu_k$ sum is constrained, such that $s$ of the $\nu_k$'s are equal to 1, and $l-s$ are equal to 2.  Similarly,
\begin{align}
    c_{n_1n_2}^{(2)} &= c_{0,1}^{(2)} \left[\sum_{l=1}^{n_1+n_2-1} \sum_{s=0}^{l} (\gamma_1)^s  (\overline{\gamma}_2)^{l-s} \prod_{k=1}^l \sum_{\substack{\nu_k,\\ i_k, j_k}} c_{i_k j_k}^{(\nu_k)} \right]. \label{eq:C2Coeff}
\end{align}

We can use these recursion relations to construct generating functions. We define
\begin{align}
    C_1(z_1, z_2, \gamma_1, \gamma_2) &\equiv \sum_{n_1, n_2} c_{n_1 n_2}^{(1)} z_1^{n_1} z_2^{n_2}, \\
    C_2(z_1, z_2, \gamma_1, \gamma_2) &\equiv \sum_{n_1, n_2} c_{n_1 n_2}^{(2)} z_1^{n_1} z_2^{n_2}.
\end{align}
Multiplying Eqs.~\eqref{eq:C1Coeff} and \eqref{eq:C2Coeff} by $z_1^{n_1} z_2^{n_2}$ and summing over $n_1$ and $n_2$ gives
\begin{align}
    C_1
    &= c_{1,0}^{(1)} z_1 \left[\sum_l \left(\sum_{ij} (\overline{\gamma}_1 c_{ij}^{(1)} + \gamma_2 c_{ij}^{(2)}) z_1^i z_2^j \right)^l \right] \nonumber\\
    &= c_{1,0}^{(1)}z_1 \sum_l (\overline{\gamma}_1 C_1 + \gamma_2 C_2)^l, \label{eq:C1Expr}\\
    C_2
    &= c_{0,1}^{(2)} z_2 \left[\sum_l \left(\sum_{ij} (\gamma_1 c_{ij}^{(1)} + \overline{\gamma}_2 c_{ij}^{(2)}) z_1^i z_2^j \right)^l \right] \nonumber \\
    &= c_{0,1}^{(2)}z_2 \sum_l (\gamma_1 C_1 + \overline{\gamma}_2 C_2)^l. \label{eq:C2Expr}
\end{align}
Hence, the generating functions form a system of equations:
\begin{align}
    C_1 &= \frac{c_{1,0}^{(1)} z_1}{1-(\overline{\gamma}_1 C_1 + \gamma_2 C_2)} = \frac{z_1}{1-(\overline{\gamma}_1 C_1 + \gamma_2 C_2)}, \\
    C_2 &= \frac{c_{0,1}^{(2)} z_2}{1-(\gamma_1 C_1 + \overline{\gamma}_2 C_2)} = \frac{z_2}{1-(\gamma_1 C_1 + \overline{\gamma}_2 C_2)}.
\end{align}

The coefficients of the generating functions $C_1$ and $C_2$ enumerate the number of diagrams that can be created with hair, with the constraints that the first step is fixed to be of type 1 or 2 and cannot be retraced until the very end. We now introduce a new function which counts the paths when we relax these constraints. To this end, it is worth noting how the constraints  enter into Eqs.~\eqref{eq:C1Expr}--\eqref{eq:C2Expr} in the first place. Taking $C_1$ as an example, the constraint of fixing the first step to be of type 1 is captured by the overall factor of $c_{1,0}^{(1)}z_1$. The constraint of forbidding this step from being retraced until the very end is captured by the reduced coordination number $\overline \gamma_1$. Thus, if we define $h_{n_1n_2}$ to be the number of ways of generating a hair that starts at a given point, with \textit{no} constraints aside from the number of steps of each type,  its generating function is
\begin{align}
    H_0(z_1, z_2, \gamma_1, \gamma_2) &\equiv \sum_{n_1, n_2} h_{n_1 n_2} z_1^{n_1} z_2^{n_2} \\
    &= \left[\sum_l \left(\sum_{ij} (\gamma_1 c_{ij}^{(1)} + \gamma_2 c_{ij}^{(2)}) z_1^i z_2^j \right)^l \right]. \label{eq:H0Expr}
\end{align}
Here $i,j$ correspond to the number of steps of each type in a given subhair, and $l$ specifies the number of subhairs. Evaluating Eq.~\eqref{eq:H0Expr}, we have
\begin{align}
    H_0(z_1, z_2, \gamma_1, \gamma_2) &= \sum_l (\gamma_1 C_1 + \gamma_2 C_2)^l \\
    &= \frac{1}{1-(\gamma_1 C_1 + \gamma_2 C_2)}.
\end{align}
We also construct $h^{(1)}_{n_1n_2}$ and $h^{(2)}_{n_1n_2}$, which correspond to the number of ways of adding a hair where the first step is not fixed, but we forbid it to be along one chosen type-1 or type-2 direction. The generating functions are similar:
\begin{align}
    H_1(z_1, z_2, \gamma_1, \gamma_2) &\equiv \sum_{n_1, n_2} h_{n_1n_2}^{(1)} z_1^{n_1} z_2^{n_2} \\
    &= \sum_l (\overline{\gamma}_1 C_1 + \gamma_2 C_2)^l = \frac{C_1}{z_1},\nonumber\\
    H_2(z_1, z_2, \gamma_1, \gamma_2) &\equiv \sum_{n_1, n_2} h_{n_1n_2}^{(2)} z_1^{n_1} z_2^{n_2}  = \frac{C_2}{z_2}.
\end{align}

To calculate $H_0$, $H_1$, and $H_2$ in practice, we must solve the system of equations
\begin{align}
    C_1 - \overline{\gamma}_1 C_1^2 - \gamma_2 C_1 C_2 &= z_1 \\
    C_2 - \overline{\gamma}_2 C_2^2 - \gamma_1 C_1 C_2  &= z_2 \\
    H_0 - \gamma_1 H_0 C_1 - \gamma_2 H_0 C_2 &= 1 \\
    H_1 z_1 -C_1 &= 0 \\
    H_2 z_2 -C_2 &= 0.
\end{align}
Generically, there is no closed form solution for this system of equations. Instead, we can write the equivalent system of equations for the coefficients of each generating function,
\begin{widetext}
\begin{align}
    c_{n_1 n_2}^{(1)} -\overline{\gamma}_1 \sum_{jk} c_{jk}^{(1)} c_{n_1 - j,n_2-k}^{(1)} - \gamma_2 \sum_{jk} c_{jk}^{(1)} c_{n_1 - j,n_2-k}^{(2)} &= \delta_{n_1,1} \delta_{n_2,0} 
    \label{eq:recursion1}
    \\
    c_{n_1 n_2}^{(2)} -\overline{\gamma}_2 \sum_{jk} c_{jk}^{(2)} c_{n_1 - j,n_2-k}^{(2)} - \gamma_1 \sum_{jk} c_{jk}^{(2)} c_{n_1 - j,n_2-k}^{(1)} &= \delta_{n_1,0} \delta_{n_2,1} 
    \label{eq:recursion2}
    \\
    h_{n_1n_2} -\gamma_1 \sum_{jk} h_{jk} c_{n_1 - j,n_2-k}^{(1)} - \gamma_2 \sum_{jk} h_{jk} c_{n_1 - j,n_2-k}^{(2)} &= \delta_{n_1,0} \delta_{n_2,0}  \\
    h_{n_1n_2}^{(1)} &= c_{n_1+1,n_2}^{(1)}  \\
    h_{n_1n_2}^{(2)} &= c_{n_1,n_2+1}^{(2)}, 
\end{align}
\end{widetext}
and solve this system of equations recursively.
Comparing Eqs.~\eqref{eq:C1Coeff}-\eqref{eq:C2Coeff} to Eqs.~\eqref{eq:recursion1}-\eqref{eq:recursion2}, we see a substantial reduction in computational effort.  Equations \eqref{eq:recursion1}-\eqref{eq:recursion2} are quadratic in the $c$'s, and calculating a new coefficient $c_{n_1n_2}$ only requires of order $n_1\times n_2$ operations.  Conversely, the time required to use the relationships in 
Eqs.~\eqref{eq:C1Coeff}-\eqref{eq:C2Coeff} grows exponentially. For the relevant case, where $\gamma_1=4$ and $\gamma_2=2$, these recursive relationships give
\begin{align}
    H_0(z_1, z_2) &= 1 + 4z_1 + 2z_2 + 28z_1^2 + 32z_1z_2 + 6z_2^2 + \dots, \label{eq:H0Series} \\
    H_1(z_1, z_2) &= 1 + 3z_1 + 2z_2 + 18z_1^2 + 26z_1z_2 + 6z_2^2 + \dots, \label{eq:H1Series} \\
    H_2(z_1, z_2) &= 1 + 4z_1 + z_2 + 28z_1^2 + 20z_1z_2 + 2z_2^2 + \dots. \label{eq:H2Series} 
\end{align}
We calculate these series up to $0\leq n_1,n_2\leq 100$.

It is also useful to consider the case where there is only one type of hopping
(such as the square lattice, $t' = 0$, or the triangular lattice, $t' = t$).  The equations are much simpler there,
\begin{align}
    C_1 - \overline{\gamma}_1 C_1^2 &= z_1 \\
    H_0 - \gamma_1 H_0 C_1 &= 1 \\
    H_1 z_1 -C_1 &= 0.
\end{align}
This system \textit{does} have a closed-form solution:
\begin{align}
    C_1 &= \frac{1 - \sqrt{1-4\overline{\gamma}_1 z_1}}{2\overline{\gamma}_1}, \\
    H_0 &= \frac{2 \overline{\gamma}_1}{2 \overline{\gamma}_1 - \gamma_1 + \gamma_1 \sqrt{1-4\overline{\gamma}_1 z_1}}, \\
    H_1 &= \frac{1 - \sqrt{1-4\overline{\gamma}_1 z_1}}{2\overline{\gamma}_1 z_1} = \frac{C_1}{z_1}.
\end{align}
For example, in the case of the square lattice ($\gamma_1=4$), we have the series expansions
\begin{align}
    H_0(z_1) &= 1 + 4z_1 + 28z_1^2 + 232z_1^3 + \dots, \\
    H_1(z_1) &= 1 + 3z_1 + 18z_1^2 + 135z_1^3 + \dots,
\end{align}
in agreement with Refs.~\cite{RaghavanPRL1995, RaghavanThesis}, and Eqs.~\eqref{eq:H0Series}--\eqref{eq:H2Series} when $z_2 \equiv 0$. Similarly, in the case of the triangular lattice ($\gamma_1 = 6$), we have
\begin{align}
    H_0(z_1) &= 1 + 6z_1 + 66z_1^2 + 876z_1^3 + \dots, \\
    H_1(z_1) &= 1 + 5z_1 + 50z_1^2 + 625z_1^3 + \dots,
\end{align}
in agreement with Eqs.~\eqref{eq:H0Series}--\eqref{eq:H2Series} when $z_1 = z_2$. 

\begin{figure}[t!]
    \centering    
    \includegraphics[width=\linewidth]{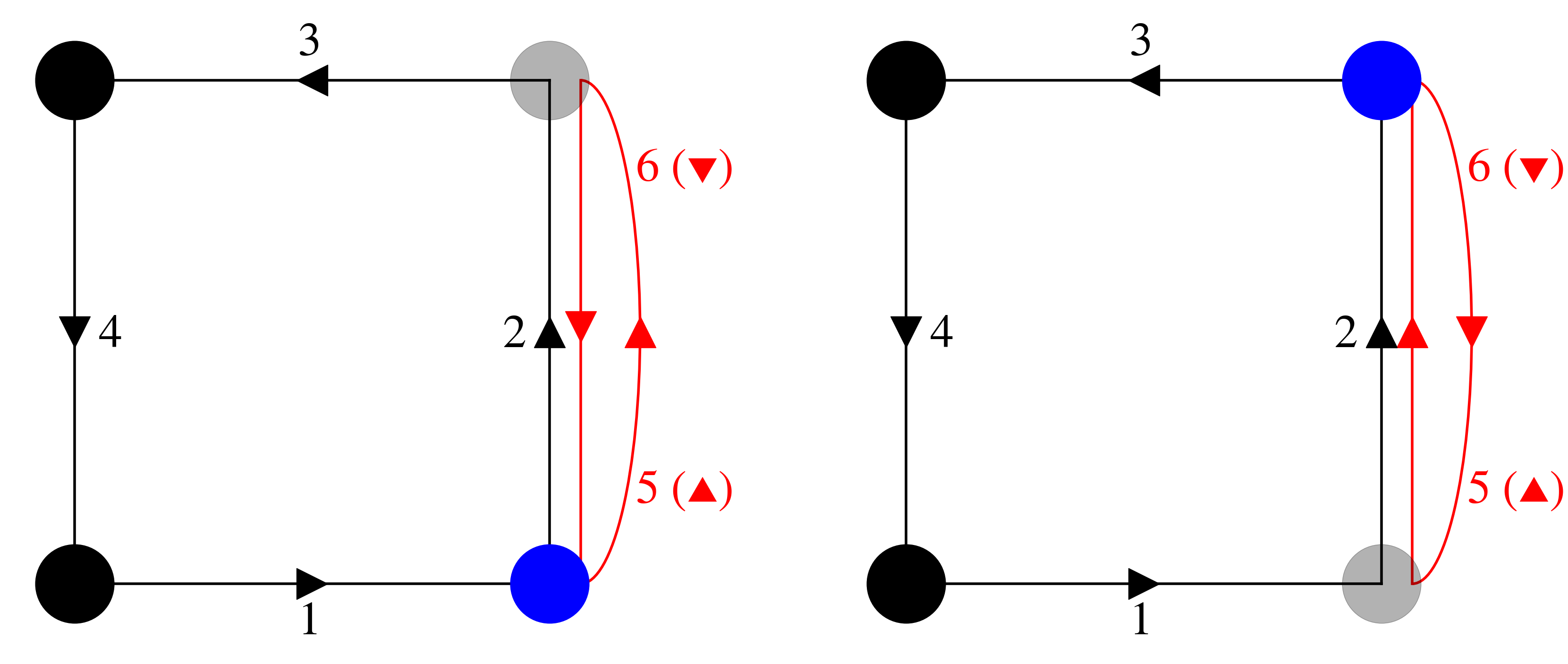}
    \caption{Two different ways of adding hair (red lines) to a hairless path (black lines). The blue vertex indicates the origin of the hair, and the gray vertex indicates the point visited by the hair. The left path is $\{1, [5, 6], 2, 3, 4\}$ = $\{\rightarrow, [\uparrow, \downarrow], \uparrow, \leftarrow, \downarrow\}$ and the right path is $\{1, 2, [6, 5], 3, 4\}$ = $\{\rightarrow, \uparrow, [\downarrow, \uparrow], \leftarrow, \downarrow\}$, with the hair indicated in square brackets. Although these paths differ in the placement of the hair, they produce the identical final path $\{\rightarrow, \uparrow, \downarrow, \uparrow, \leftarrow, \downarrow\}$ and hence double-count the same diagram in the high-temperature expansion. To prevent such over-counting, we forbid a hair from starting along the edge that preceded it, such as the diagram on the right where step 6 $(\downarrow)$ backtracks step 2 $(\uparrow)$.}
    \label{fig:HairOvercount}
\end{figure}

We now show how to construct $w(\ell_p,\ell_p^\prime,n_1,n_2)$ from these coefficients. For illustrative purposes, we assume $t'=0$ and consider the square lattice where $\ell_{p}^\prime=n_2=0$. Figure~\ref{fig:Hair} shows an exemplary hairless closed path, a square plaquette. The generating function for the weight function coefficients $w$ of this diagram is $Q(\ell_p=4)=\sum_{n_1} w(\ell_p=4,0,n_1,0) z^{n_1}$. There are five vertices in this diagram (the origin being counted twice, as the starting and ending location of the dopant). Thus we can consider adding arbitrary hairs at each of these five vertices.  Unfortunately, as illustrated in Fig.~\ref{fig:HairOvercount}, adding certain hairs to neighboring vertices can produce identical diagrams.  This double-counting can be avoided by forbidding the first step in any subhair from pointing along the edge that precedes the selected vertex.  This constraint is captured by the generating function $H_1$ and applies to all vertices except the first, the origin. Hairs starting from the origin are unconstrained, as the origin has no preceding edge. These hairs are thus counted using the generating function $H_0$. Enumerating all possible ways of adding hairs to the square plaquette therefore yields $Q=H_0 H_1^4$. Raghavan and Elser erroneously used $Q=H_0^5$ in Refs.~\cite{RaghavanPRL1995, RaghavanThesis}. For generic diagrams our argument clearly generalizes to $Q=H_0 H_1^{\ell_p}H_2^{\ell_{p}^\prime}$.  We tabulate the coefficients of these generating functions for $0\leq \ell_p+\ell_{p}^\prime \leq 100$.  

{We perform two checks for correctness. Firstly, we calculate the energy and spin-spin correlations of a particle dopant using a brute force enumeration of all diagrams up to a maximum length. Secondly, we perform Monte Carlo simulations for particle and hole dopants where the addition or removal of hairs are instead sampled as another move set, as in Fig.~\ref{fig:Moves}. In both cases, there is excellent agreement with our resummed approach, although the resummed simulations converge more quickly and can attain lower temperatures.}

\section{Technical Details of the Monte Carlo Simulations}
\label{subsec:QMCAppendix}

{In this Appendix, we describe some of the technical details of our Markov chain Monte Carlo simulations. 
At each temperature in our simulation, we first begin the Markov chain with $N_{\rm eq} \approx 10^9$ equilibration sweeps, during which no samples are measured. We then take $M = 10^5 \times N_{\rm bins}$ samples, with $N_{\rm bins} = 24$ and the samples taken once every $N_{\rm sw} = 3 \times 10^3$ sweeps, before cooling down to the next temperature. Equation~\eqref{eq:SampleAvg} is used to estimate quantities in a given bin, and we then average over all $N_{\rm bins}$ to produce the final result. The error is given by the standard error of the mean across all bins.}

{To confirm that the samples are uncorrelated, we perform a rebinning analysis: we combine the data in every two or three bins (such that $N_{\rm bins} = 12$ or $N_{\rm bins} = 8$, respectively) and recalculate all quantities of interest using the rebinned data. We find that the error estimates do not depend on the choice of bins.

{For our simulations, we take about 40 temperatures across two decades, from $T/\abs{t} = 10^{0.5} \approx 3.2$ to $T/\abs{t} = 10^{-0.5} \approx 0.32$, equally spaced on a logarithmic scale. As mentioned in Appendix~\ref{subsec:HairAppendix}, we tabulate hairs with $0 \leq n_1, n_2 \leq 100$ for the resummation of the series. For the move probabilities, we choose $p_{\rm plaq}^\square = p_{\rm plaq}^\triangle = 0.025$ (Fig.~\ref{fig:Moves}(a) and Fig.~\ref{fig:Moves}(b)), $p_{\rm edge}^\square = p_{\rm edge}^\triangle = 0.35$ (Fig.~\ref{fig:Moves}(c) and Fig.~\ref{fig:Moves}(d)) and $p_{\rm corner} = 0.25$ (Fig.~\ref{fig:Moves}(e,f)). Note that corner moves are treated as a single move class. This is because plaquette and edge moves explicitly make reference to square or triangular plaquettes and are therefore divided into two classes. Corner moves, on the other hand, just swap two successive edges regardless of the underlying geometry.}

\bibliography{refs.bib}

\end{document}